\documentclass[aps,pre,twocolumn,groupedaddress,superscriptaddress,showpacs]{revtex4-1}
\usepackage{tikz}
\usepackage{amsmath}
\usepackage{graphicx}
\usepackage{float}
\usepackage{color}
\usepackage{tikz}
\usepackage{float}
\usepackage{fancyvrb}

\begin{document}

	\title{ Molecular Dynamics Simulations of Active Matter using LAMMPS} 

    \author{C. S.  Dias}
    \email{csdias@fc.ul.pt}
    \affiliation{Departamento de F\'{\i}sica, Faculdade de Ci\^{e}ncias, Universidade de Lisboa, 
    1749-016 Lisboa, Portugal}
    \affiliation{Centro de F\'{i}sica Te\'{o}rica e Computacional, Faculdade de Ci\^{e}ncias, Universidade de Lisboa, 1749-016 Lisboa, Portugal}

\begin{abstract}

LAMMPS is a widely popular classical Molecular Dynamics package. It was designed for materials modeling but it is well prepared for simulations in Soft Matter.  The use packages like LAMMPS has advantages and disadvantages. The main advantage is the optimization of the methods, mainly for parallel computing. The main disadvantage is that, due to the complexity of the code, it ha long learning curve. One purpose of these notes is to shorten that curve for researchers are starting to use LAMMPS to simulate soft active matter. In these notes, we first discuss some Molecular Dynamics methods implemented in LAMMPS. We present an hands-on introduction to first-time users and we finish with an advanced hands-on section, where we implement and test Active Brownian particles simulations.
  
\textit{Keywords}: LAMMPS,  Molecular Dynamics, Langevin Dynamics, Active Matter, Active Brownian particles

\end{abstract}

\maketitle
 
\section{Introduction}

LAMMPS is an acronym for Large-scale Atomic/Molecular Massively Parallel Simulator. It is a classical Molecular Dynamics package optimize for parallelization and mainly directed for materials modeling. Its development began in the mid 1990s, with its first versions in FORTRAN 77 and 90. It was build under a cooperative research \& development agreement (CRADA) between two DOE labs (Sandia and LLNL) and 3 companies (Cray, Bristol Myers Squibb, and Dupont).  The coding effort was led by Steve Plimpton at Sandia \cite{IntVeld2008,Parks2008,Brown2011,Brown2012,Petersen2010,Plimpton1995}.
 
After the initial development in FORTRAN, the current versions are written in C++ and it was released as an open source code in 2004. The package is freely available for download under GPL, and is designed to be easy to modify or extend. The package is well documented (see \url{https://lammps.sandia.gov}). 
 
LAMMPS is a very powerful package for Molecular Dynamics simulations, and the use of this type of package can have many advantages and disadvantages. The advantages are clear, since it is designed to be modular, it is relatively easy to implement a new model and is optimize for parallel computing. The use of this type of packages gives us a confidence in the code since it is tested by many. In the other hand, it has also a few disadvantages.  For instance, also because it is modular, the implemented routines are thought-out to be generic, which lead to a lower performance than methods implemented for specific models.  One main disadvantage is also the complexity of the code, since it is difficult to follow everything implemented in the code, it has a long initial learning curve. The objective of these notes is to shorten that initial learning curve for researchers starting to simulate soft active matter with LAMMPS.
 
These notes are divided into three main sections. The first section discusses basic Molecular Dynamics methods behind LAMMPS and their implementation. The second section is a hands-on introduction to LAMMPS. The third section is a more advanced hands-on section where we implement a custom method for the simulation of Active Brownian particles.

\section{Basic Molecular Dynamics in LAMMPS}\label{sec:basic}

In this section we discuss the basic Molecular Dynamics methods implemented in LAMMPS. Since Molecular Dynamics simulations consist in integrating the equations of Newton for many particles \cite{Allen87}, we start by discussing the integration method.  This method needs the force computation, that is discussed in the following section. Some concepts of parallelization are discussed next. We then show how Langevin Dynamics simulations are performed in LAMMPS, and the last section has a brief discussion on reduced units.

\subsection{Velocity Verlet integration}\label{sec:verlet}

Many different methods can be used to integrate the equations of Newton. Here we discuss the velocity Verlet method.
This method is similar to the classical Verlet method and consists of two integration steps \cite{Swope1982}:
\begin{equation}
\vec{r}(t+\Delta t)=\vec{r}(t)+\vec{v}(t)\Delta t+\frac{1}{2}\vec{a}(t)\Delta t^2, \label{eq:vel_verlet_1} 
\end{equation}
and, 
\begin{equation}
\vec{v}(t+\Delta t)=\vec{v}(t)+\frac{\vec{a}(t)+\vec{a}(t+\Delta t)}{2}\Delta t, \label{eq:vel_verlet_2}
\end{equation}
where $\vec{r}$ is the position,  $\vec{v}$ the velocity, and $\vec{a}$ the acceleration. $\Delta t$ is the integration timestep. One advantage of the velocity Verlet method when compared to the classical Verlet one is that it deals directly with the velocity,  which avoid problems in the definition of the initial velocity,. 

\textbf{LAMMPS implementation:} One could implement directly the two steps given by Eqs.~\ref{eq:vel_verlet_1}~and~\ref{eq:vel_verlet_2}, however these steps need to save, at all time, two acceleration (or force) vectors, which take up memory and communication time between parallel jobs. The most common implementation, and the one used in LAMMPS, computes the velocity twice, using an intermediate half timestep jump forward. The steps are the following: 
\begin{enumerate}
\item Compute $\vec{v}(t+\Delta t/2)=\vec{v}(t)+\vec{a}(t)\frac{\Delta t}{2}$; 
\item Compute $\vec{r}(t+\Delta t)=\vec{r}(t)+\vec{v}(t+\Delta t/2)\Delta t$;
\item Compute the forces at $\vec{r}(t+\Delta t)$; 
\item Compute $\vec{v}(t+\Delta t)=\vec{v}(t+\Delta t/2)+\frac{1}{2}\vec{a}(t+\Delta t)\Delta t$. 
\end{enumerate}
With this implementation, only one vector of forces is used since they computed between the two steps.

\subsection{Force computation}\label{sec:force}

\begin{figure}[t]
    \centering
    \includegraphics[width=\columnwidth]{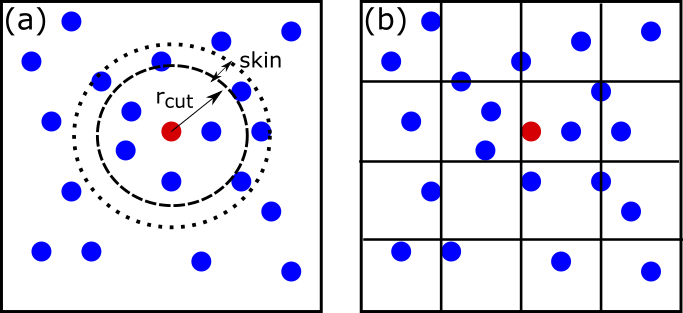}
    \caption{(a) Neighbor list (Verlet list) used to choose the pairs of particles to compute the force. $r_{cut}$ is the cutoff distance of the potential and $skin$ is an extra distance added to $r_{cut}$ as a threshold to build the list of neighbors. (b) Cell list method where the system is divided into cells. In LAMMPS it is not used to compute the force but to build the neighbor list without searching all particles.}
    \label{fig:neighblist}
\end{figure}

The force calculation is usually the bottleneck for an efficient Molecular Dynamics simulation. For the simplest case of pair potentials, for instance, the calculation is done over all pairs of particles. One first optimization is by considering that the potential interaction between particles can be neglected above a certain distance, $r_{cut}$.  Taking that into account, pairs of particles at distances above $r_{cut}$ do not need to be computed. Different methods can be used to take advantage of that. The most common ones are the Verlet list (see Fig.~\ref{fig:neighblist}(a)) and the cell list (see Fig.~\ref{fig:neighblist}(b)) methods. The first consists on saving a list of neighbor particles and update the list from time to time. The second is to discretize the system into cells and only check particles in the neighboring cells.

\textbf{LAMMPS implementation:} In LAMMPS, both methods are implemented.The calculation of the force is done using the Verlet list (see Fig.~\ref{fig:neighblist}(a)). The list is defined as all particles that are at a distance of $r_{cut}+skin$, where the $r_{cut}$ depends on the potential and the $skin$ is defined (command ``neighbor" in the LAMMPS script). The cell list method (see Fig.~\ref{fig:neighblist}(b)) is then used to update the particles in the Verlet list, otherwise it would be necessary to run all pairs of particles. The length of the cell can be defined and it is a trade-off between the time it takes to run all cells if the cells are too small, or the number of unnecessary particles check if they are two large. LAMMPS consider the length of the cell $r_{cut}/2$ as default but it can be controlled (command ``neigh\_modify" in the LAMMPS script).

\subsection{Parallel computing} \label{sec_parallel}

\begin{figure}[t]
    \centering
    \includegraphics[width=0.7\columnwidth]{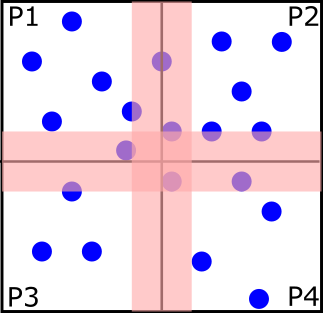}
    \caption{Spacial domain decomposition by processes 1 to 4. Particles are allocated to each process and particles in the red region that belong to a process are ghost particles in the process of the adjacent domain.}
    \label{fig:parallel}
\end{figure}

One of the main advantages of Molecular Dynamics simulations is its scalability when split into multiple processes.  Basically, most tasks can be done in parallel, needing only information from the positions of other particles during the force calculation. One of the most common method for parallel computing is the spatial decomposition method \cite{Plimpton1995}. As you can see in Fig.~\ref{fig:parallel}, the space is divided into domains that are assigned to different processes.  Particles inside the  red region that belong to a certain process are also ghost particles to processes in adjacent domains. This way, when the force is computed, all particles in the domain plus the ghost particles are used.

\textbf{LAMMPS implementation:} In LAMMPS, the length of the red region is given by $r_{cut}+skin$ (the command ``comm\_modify" in the LAMMPS script can be used to change the length of the red region). The mapping of the spacial domains can be controlled in LAMMPS (command ``processors" in the LAMMPS script), however, by default, LAMMPS optimize the domain distribution in such a way to minimize the red region volume.  In the Molecular Dynamics pipeline one needs to add a step where the communication between processes is done and the particles that belong to each process are updated. In LAMMPS, this communication step occurs after the first integration of the velocity Verlet method (before step 3 of force computation in Sec.~\ref{sec:verlet}).

\subsection{Langevin dynamics}\label{sec:langevin}

Soft matter dynamics usually occur inside viscous environments (fluid),  so to perform Molecular Dynamics simulations one needs to add the interaction with the fluid. It is possible to add the particles of the fluid and perform brute force simulations, however the length and time scales will be very short. One approximation is to coarse-grain the interaction with the  fluid using the Langevin dynamics
approach. Langevin dynamics follows the Langevin equation where two new terms of force are added: viscous drag and thermal noise. The Langevin equation for the translational degrees of freedom, in one dimension, is given by,
  \begin{equation}
   m\dot{v}(t)=-\nabla_{x} V(x)-\gamma_t v(t)+\xi_t(t), \label{eq.Langevin}
  \end{equation}
where $m$ is the mass of colloids, $\gamma_t$ the translational damping coefficient related to a drag force
exerted by the medium, $V$ the external potential, and $\xi_t(t)$ is the stochastic term, 
from the thermal noise, given from a distribution of zero mean and satisfying the relation, 
\begin{equation}
 <\xi_t(t)\xi_t(t')>=2k_BT\gamma_t \delta(t-t'), \label{eq.langevin_noise}
\end{equation}
where $\delta$ is the Dirac delta,  $T$ is the temperature, $k_B$ the Boltzmann constant.
For a rigid body, one also needs the Langevin equation for the rotational degrees of freedom. The Langevin equation for the one-dimensional diffusion of a single-axis rotation is given by,
  \begin{equation}
   I\dot{w}(t)=-\nabla_{\theta} V(\theta)-\gamma_r w(t)+\xi_r(t), \label{eq.Langevin_rot}
  \end{equation} 
where $I$ is the moment of inertia and  $\gamma_r$ the rotational damping coefficient along that single-axis rotation, $\xi_r(t)$ is the stochastic torque
from the thermal noise, given from a distribution of zero mean and satisfying the same relation of, 
\begin{equation}
 <\xi_r(t)\xi_r(t')>=2k_BT\gamma_r \delta(t-t').\label{eq.langevin_noise_rot}
\end{equation}
For both translational and rotational motion, above a damping time (time for the inertia to become negligible), the Brownian diffusive regime is recovered. We obtain the translational diffusion coefficient given by the Stokes-Einstein relation as \cite{Mazza2007},
 \begin{equation}
 D_t=\frac{k_BT}{\gamma_t}, \label{eq.SE_gen}
\end{equation}
and the rotational diffusion coefficient as,
 \begin{equation}
 D_r=\frac{k_BT}{\gamma_r}. \label{eq.SED_gen}
\end{equation}
The damping times for the translational and rotational motions are $\tau_t=m/\gamma_t$ and $\tau_r=I/\gamma_r$, respectively. 

The translational and rotational diffusion coefficients presented here are generic for passive or active particle. In the case of active particles, the rotational diffusion coefficient sets the time scale for the self-propulsion directional motion to forget its initial orientation (for active particles the effective diffusion coefficient depends also on the propulsion which is discussed in Sec.\ref{sec:adv_hand}). For passive particles, one can find a relation between translational and rotational diffusion. For spherical passive particles,  these coefficients take the form \cite{Mazza2007},
 \begin{equation}
 D_t=\frac{k_BT}{6\pi\eta R}, \label{eq.SE}
\end{equation}
and,
 \begin{equation}
 D_r=\frac{k_BT}{8\pi\eta R^3},\label{eq.SED}
\end{equation}
where $\eta$ is the viscosity of the fluid medium and $R$ the radius of the particles.
It is possible to obtain a relation between the rotational and translational diffusion coefficient given by,
\begin{equation}
 \frac{D_r}{D_t}=\frac{3}{4R^2}. \label{eq.coeff_rel}
\end{equation}

\textbf{LAMMPS implementation:} The variables, in LAMMPS, that parameterize the Langevin dynamics are the translational damping time, $\tau_t$ and the mass $m$ (command ``fix langevin" in the LAMMPS script).  Since only the translational damping time is defined, an isotropic scale factor $\alpha=\tau_t/\tau_r$, between translational and rotational damping, is introduced.  For spherical particles, using Eq.~\ref{eq.coeff_rel}, one can write the scale factor as $\alpha=10/3$.  For rigid aspherical particles, $\alpha$ is an external parameter (``angmom" in the langevin fix), and for rigid more complex bodies, if they are approximately spherical you need to add the factor of 10/3 directly into the code \cite{Dias2016}, or for more complex shapes the scale factor is one \cite{Melo2020}. In LAMMPS, the Langevin equation for the translational motion, in one dimension, is given by, 
\begin{equation}
 m\dot{v}(t)=-\nabla_r V(r)-\frac{m}{\tau_t}v(t)+\sqrt{\frac{2mk_BT}{\tau_t}}\xi(t), \label{eq.trans_Langevin_dynamics}
\end{equation}
and for the rotational motion, around one single axis of rotation,  is,
\begin{equation}
 I\dot{\omega}(t)=-\nabla_{\theta} V(\theta)-\frac{\alpha I}{\tau_t}\omega(t)+\sqrt{\frac{2\alpha Ik_BT}{\tau_t}}\xi(t).\label{eq.rot_Langevin_dynamics}
\end{equation}
In eqs.~\ref{eq.trans_Langevin_dynamics}~and~\ref{eq.rot_Langevin_dynamics}, $v$ and $\omega$ 
are the translational and angular velocity, $m$ and $I$ are
mass and Inertia of the colloidal particle, and $V$ is the external potential. In LAMMPS, for efficiency \cite{Dunweg1991}, $\xi(t)$ is an uniform
distribution from $-0.5$ to $0.5$, which forces a correction in the stochastic term of the 
Langevin equation since the second moment of the uniform distribution is given by,
\begin{equation}
 \sigma^2=\int^{0.5}_{-0.5}x^2dx=\frac{2}{3}\left(\frac{1}{2}\right)^3=\frac{2}{24}, \label{eq.uniform_sigma}
\end{equation}
and imposes a correction of $\sqrt{2/24}$,  that can be seen in the source code in the file fix\_langevin.cpp.
In the same source file, the integration timestep is included in the denominator inside the square root
because LAMMPS multiply all forces (including the stochastic term from the thermal
noise) by the timestep in the velocity Verlet integration (see Sec.~\ref{sec:verlet}). The 
timestep in the stochastic term comes from the variance of the random number distribution
which is $\sigma^2=2D\Delta t$ (which is the multiplicative factor in the stochastic term of the Langevin equation). This way, the diffusion coefficient does not depend on the timestep.

\subsection{Reduced Units}

When modeling Soft Matter, we are usually working with length scales in the order of nanometers to micrometers and energy scales in the order of $10^{-1}$ to $10^{-2}$ eV.  If we decide to use standard SI units, we will have to work with very small values.  It is, of course, more convenient to work with values near unity, which can be done by using the size of the simulation elementary particles as the reference. Reduced units are not only used for convenience. Performing numerical calculations with very small values can lead to rounding errors due to the machine precision.Using dimensionless units also facilitates scaling, where a single model can help to understand a large class of problems at different scales \cite{Rapaport1996}.

\textbf{LAMMPS implementation:} A common scale for reduced units is the one of the Lennard-Jones potential \cite{Rapaport1996}, given by,
\begin{equation}
V_{LJ}(r)=4\epsilon\left[\left(\frac{\sigma}{r}\right)^{12}-\left(\frac{\sigma}{r}\right)^6\right],
\end{equation}
where $r$ is the distance between two particles, $\epsilon$ the depth of the potential well, and $\sigma$ the width of the potential (distance at which the potential is zero). 
LAMMPS uses these units and, without loss of generality, sets the fundamental quantities $m$, $\sigma$, $\epsilon$, and $k_B$ to unity. 
The masses, distances, energies are multiples of these fundamental values. 
The formulas relating the reduced or unitless quantity are show in Table~\ref{table.red_units}.

\begin{table}[h]
\begin{center}
\caption{LAMMPS reduced units.}
\begin{tabular}{cc}
\hline
Variable & Value \\
\hline
Temperature ($T$) & $\frac{\epsilon}{k_B}$ \\
Time ($\tau$) & $\sqrt{\frac{m\sigma^2}{\epsilon}}$ \\
Boltzmann constant ($k_B$) & $1$ \\
Energy ($\epsilon$) & $1$ \\
Distance ($\sigma$) & $1$ \\
Mass ($m$) & $1$ \\
Translational damping time ($\tau_t$) & $\sqrt{\frac{m\sigma^2}{\epsilon}}$ \\
\hline
\end{tabular}
\label{table.red_units}
\end{center}
\end{table}

The conversion to real units can be sometimes confusing. Nonetheless, it is always better to use reduced units instead of SI units (which LAMMPS allows). The conversion between real units and reduced units can be done by using the relations in Table~\ref{table.red_units}. For instance, Soft Matter experiments are usually performed at room temperature of $T=293K$. So taking Boltzmann constant, $k_B=1.38\times10^{-23}m^2kgs^{-2}K^{-1}$, we obtain the energy unit of $\epsilon=4.04\times10^{-21}J$.  The same can be done with the length units,  for instance, if we set the mass of a colloid as $M=10m$ and the diameter as $d=10\sigma$ in the simulations, and consider the real values of colloid mass $10^{-12}g$ and radius $0.5\mathrm{\mu m}$, we recover the fundamental quantities of $\sigma=10^{-7}m$ and $m=10^{-16}kg$ respectively.  The same can be done with the time units. If we set a timestep in the simulations of $\Delta t=0.01$, this tells us that each simulation step takes $1.57\times 10^{-7}s$. For Langevin Dynamics simulations it is sometimes clearer to use the Brownian time \cite{Dias2018c} (time for a particle to diffuse the square of its diameter) to set the timescale.

\section{LAMMPS: basic hands-on}\label{sec:bas_hand}

In this section we will have a more practical approach to LAMMPS. We will start by downloading and compile LAMMPS, run a test example, and performing some measurements on a Langevin Dynamics simulation of passive particles.

\subsection{Compiling LAMMPS}

LAMMPS was primarily developed for Linux systems, mainly because most of the high-performance computers are Linux based. With this in mind, if you have a Linux system it is straightforward to download the most recent tarball (\url{https://lammps.sandia.gov/download.html}), uncompress and compile performing ``make mpi" or ``make serial" in the source folder (src/). If you have a windows 10 or newer I recommend to use the Windows Subsystem for Linux (WSL). Clear instructions are available at \url{https://lammps.sandia.gov/doc/Howto_wsl.html}. For older versions of windows I recommend you to install some Linux virtual machine. MacOS users can install like in the Linux case.

In LAMMPS one can find extra packages to include some specific features. These packages were added by the LAMMPS developers or by other users, which are consistent with the style  of the rest of LAMMPS. To check what packages are installed just run the command ``make ps" in the source folder. In this course we will use, for instance, the ASPHERE package and the COLLOID package. You can add them by running the commands ``make yes-asphere" and ``make yes-colloid". You need to recompile after adding a new package.

\subsection{Running LAMMPS}

When compiling the code with the ``make mpi" or ``make serial", an executable file is created in the source folder with name lmp\_mpi or lmp\_serial. To run, one just need to use the command:
\begin{verbatim}
./lmp_mpi -in lj.in
\end{verbatim}
to run in one core, or
\begin{verbatim}
mpirun -np 4 ./lmp_mpi -in lj.in
\end{verbatim}
to run in four cores (the lmp\_serial executable do not allow parallelization). The lj.in file is a LAMMPS script where the simulation parameters and methods are included. An example of the script for Lennard-Jones interacting particles in a box with periodic boundary conditions is the following:
\begin{Verbatim}[numbers=left,xleftmargin=5mm]
units lj
atom_style atomic

dimension 3
boundary p p p
region box block 0 30 0 30 0 30 units box
create_box 1 box

lattice sc 0.1
create_atoms 1 box
mass 1 1.0
velocity all create 1 176217 dist gaussian

pair_style      lj/cut 2.5
pair_coeff      1 1  1 1

dump    DUMPXYZ     all       xyz 5 out.xyz

timestep 0.005
fix 1 all nve

thermo_style custom step temp pe ke etotal
thermo 100

run 5000
\end{Verbatim}

Lines 1 and 2 define main properties of the simulations like the reduced (LJ) units and the atom style ``atomic"defines particles as point particles. The lines 4 to 7 define and create the simulation box in three dimensions with periodic boundary conditions in all directions (boundary p p p). A region named ``box" is defined in line 6 and the simulation box with the properties of that region is created. Lines 9 to 10 create the particles in the region ``box". They are distributed on a square cube (sc) lattice with a reduced density of particles of $\rho^*=0.1$, where $\rho^*=N/V$ with $N$ the number of particles and $V=L_x\times L_y\times L_z$ the volume in reduced units.  The command ``mass" sets the mass of a particle of type 1 as 1 and the command ``velocity" sets their velocities with a Gaussian distribution around 0 and a variance equal to the temperature 1. The value 176217 is the random number generator seed.

\begin{figure}[t]
    \centering
    \includegraphics[width=1\columnwidth]{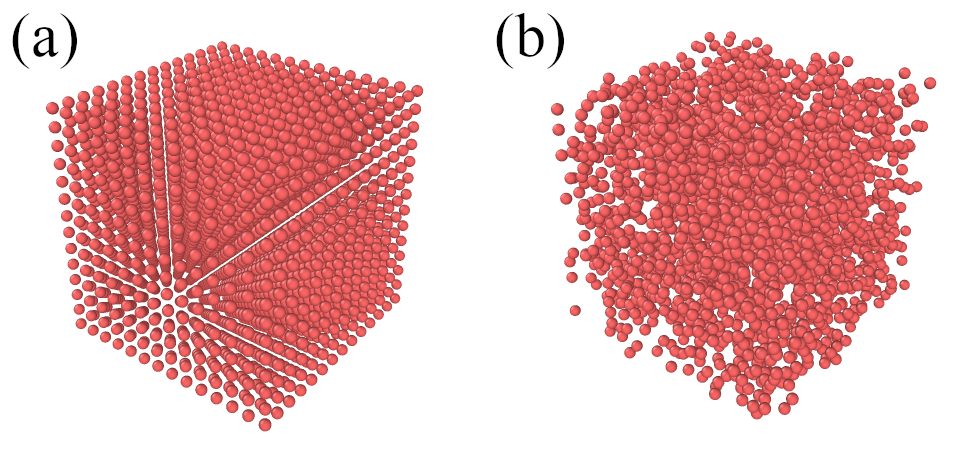}
    \caption{(a) Initial and (b) final configuration  of the NVE simulation using the test script. The visualization was performed using Ovito.}
    \label{fig:nve_configs}
\end{figure}

In line 14 and 15 the Lennard-Jones potential is defined, with a cutoff radius of 2.5 and coefficients of $\sigma=1$ and $\epsilon=1$ (two last values). The first two values of the coefficients indicates which pair of particles type will have those coefficients.  Different particles could have different interactions defined. In 17 a xyz file is created with the positions of all the particles every 5 time steps. These type of files can be read in different external software, which I recommend Ovito (\url{https://www.ovito.org}) to be simple to download and use, and very powerful.  In Fig.~\ref{fig:nve_configs}, the initial and final configuration of a simulation, using the proposed script, are shown.

In line 19 we define the simulation timestep. One can test larger timesteps to check LAMMPS error outputs when the simulation is unstable. In line 20, the fix nve is applied to all particles and perform integration of the equations of motion in constant number of atoms (N), volume (V), and energy (E). The ``fix" commands in LAMMPS are used to change something in the integration pipeline. In lines 22 and 23 the output information is defined (if no file is defined the outputs are saved in the lammps log file). In this case: timestep, temperature, potential energy, kinetic energy, and total energy. One simple test is to check the constant total energy of the NVE integration. Line 25 defines the total number of timesteps in the simulation. 

\subsection{Langevin Dynamics of passive particles in LAMMPS}

To test the concepts introduced in Sec.~\ref{sec:langevin}, we perform Langevin Dynamics simulations in LAMMPS using the following script:
\begin{Verbatim}[numbers=left,xleftmargin=5mm]
units lj
atom_style sphere

variable temperature equal 1
variable damping equal 1

dimension 3
boundary p p p

read_data       inp.data

dump    DUMPXYZ     all       xyz 50 out.xyz

timestep 0.005
fix 1 all nve
fix 2 all langevin ${temperature} &
    ${temperature} ${damping} 82763871 

compute msd_col all msd

thermo_style custom step c_msd_col[1] &
   c_msd_col[2] c_msd_col[3] c_msd_col[4] ke

thermo 100
run 50000

\end{Verbatim}

Some commands are the same as the previous script and the only relevant to Langevin Dynamics is the command in line 16 where the Langevin fix is added to the NVE integration. In this fix, basically, extra terms are added to the forces and torque following Eqs.~\ref{eq.trans_Langevin_dynamics}~and~\ref{eq.rot_Langevin_dynamics}. We add the starting and stopping temperature (which, in this case, we defined constant over the full simulation run), and the Damping time introduced in Sec.~\ref{sec:langevin}. The last value is the random number generator seed.

In this script we introduced a new atom style in line 2, which is the sphere type, where particles are considered as spheres, with a specific diameter and density. We also introduced the concept of variables in line 4 and 5 which we use in the Langevin fix. One major difference in this script is in line 10 where we read the particles positions from an external file (inp.data) with the following structure:
\begin{Verbatim}[numbers=left,xleftmargin=5mm]
# lammps molecular data

1 atoms

1 atom types

0 200 xlo xhi
0 200 ylo yhi
0 200 zlo zhi


Atoms

1  1 1 1.9099 100 100 100
\end{Verbatim}

We define the number of particles and particle types in line 3 and 5, and the size of the simulation box in lines 7, 8, and 9. The particles positions are defined in the Atoms sections, where we can see the first (and only) particle (column 1 gives then particle number) of type one (column 2), with a diameter of one (column 3), a particle density of 1.9099 (column 4) and positions in  x, y, and z (columns 5, 6, and 7). The order of the particles properties is specific for the sphere atom style, the manual of the read\_data command has examples for other  atom styles.  The density was chosen such that a particle of unit diameter has unit mass.

The final new command introduced in the run script is in line 19. The command ``compute" is used by LAMMPS to perform calculations during the simulation. Here, we compute the mean square displacement (msd) of all particles (only one in this data file). The name of the compute is msd\_col, which is used in the output ``thermo\_style" (line 21) , which is a vector with four coordinates (x, y, z, and total), summed and averaged over all particles.	

\begin{figure}[t]
    \centering
    \includegraphics[width=1\columnwidth]{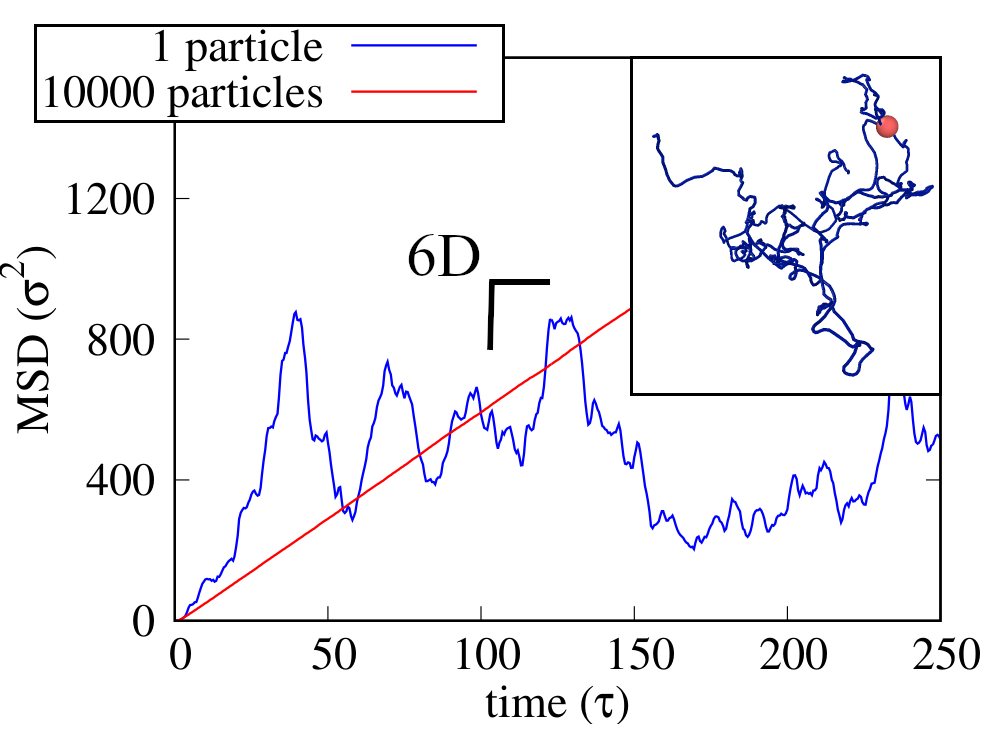}
    \caption{Mean square displacement as a function of time for a single particle and for 10000 particles. The inset shows the trajectory of a single particle.}
    \label{fig:langevin}
\end{figure}

In Fig.~\ref{fig:langevin} we plot the output of the proposed script for a single particle and for 10000 particles. To implement the many particles case, just add 10000 consecutive lines in the inp.data file. Since no pair potential has been defined one can just define the same position for all particles.  The proper slope of the mean square displacement, in three dimensions, of $6D$ is recovered, where $D$ is the diffusion coefficient of spherical passive particles given by the Stokes-Einstein relation of Eq.~\ref{eq.SE}.

\section{Implementation of Active Brownian particles in LAMMPS}\label{sec:adv_hand}

As stated before, LAMMPS is an open source code under GPL. This means you are free to download, use, study, and even change. At this point, more than 95\% of LAMMPS is made of add-on files to the core ones. Despite being a very complete and powerful tool, some specific cases are not implemented and a user could need to add new features. 

One feature, relevant for our topic of active matter, is the inclusion of Active Brownian particles simulations, i.e., particles that have a persistent motion in a specific direction. This feature has been recently implemented in LAMMPS (see command ``fix propel/self"), and others have used custom implementations \cite{Stenhammar2014}. However, in this section we discuss how to build a new class for Active Brownian particles, which can be used as an example on how to implement custom methods into LAMMPS. The implementation of Active Brownian particles consists on adding a force $\vec{F_A}=F\vec{e}(t)$ \cite{Shaebani2020,Volpe2014} where $F$ is the magnitude of the activity (a constant) and $\vec{e}(t)$ is the unit orientation vector. The orientation vector evolution over time is given by Eq. ~\ref{eq.rot_Langevin_dynamics}.  

Most of the features to be added to LAMMPS require writing a new C++ class. The simplest way is to find some class that does something similar to the one we want to add.  For instance, in the case of an active particle, one just need to add a force to a specific preferred direction of the particle. Since this is not a pair interaction force, we should not build a new pair class. In this case, we just add a force to the ones already computed in the pair classes much like is done in the Langevin dynamics case. The simplest classes one could use as examples are the fix\_addforce or the fix\_setforce, which basically add a value or set a value to the force array in a specific function of the class, that LAMMPS uses for those cases, called post\_force. The only thing we need to worry is how to define the orientation of the active particles. One option is to use ellipsoid particles (which we can define all axis of equal size if we want spheres), which then integrates the rotational motion through the use of quaternions \cite{Rapaport1996}.  In the case of ellipsoids, LAMMPS defines the x-axis as the reference axis and the rotation is defined by the quaternions values using that same axis as a reference.  You just need to have a fail check on your new build class that your atom\_style is ellipsoid.

For a fast implementation of active particles in LAMMPS one could do a simpler task. This task can be resumed into four steps:
\begin{itemize}
\item Copy the fix\_langevin class to a fix\_langevin\_active one (each class has two files, in this case fix\_langevin.cpp and fix\_langevin.h). You just need to change the class name inside the files, and in fix\_langevin\_active.h change the call of the fix from ``langevin" to ``langevin/active" for instance. This way, we use a fix, which is a bit more complex to fully understand than the two examples used before,  but is far easier to implement. For starter, the use of ellipsoids is already contemplated. 
\item In the constructor function, in the beginning of the class, we can access the arguments of the fix. An option for ellipsoids is the "angmom" (which is the scale factor $\alpha$ of the rotational diffusion in Eq.~\ref{eq.rot_Langevin_dynamics}), which has a numerical value after the option. We only need to add a second value with the force (just copy the line with the variable ``ascale" and add your new variable from ``arg[iarg+2]") that represents the activity, which will be saved into a variable defined in the header file. 
\item The final step to have a fully functioning class with activity is to add, in the angmom\_thermostat() function, a few lines inside the cycle that runs over the local particles (particles in the same process) which will add a force to the forces vector, in the direction given by the ellipsoid quaternions. One can convert quaternions to the three main axis of the ellipsoids using a function in MathExtra called q\_to\_exyz (use only the principal axis of the ellipsoid). 
\end{itemize}

After implementing this new fix, we just need to recompile LAMMPS with the new files in the source folder. We then run a very similar script as the one for the Langevin dynamics,
\begin{Verbatim}[commandchars=\\\{\}, numbers=left,xleftmargin=5mm]
units lj
atom_style ellipsoid

variable t equal 0.1
variable d equal 0.1

dimension 3
boundary p p p

read_data       inp.data

compute shape all property/atom shapex shapey &
                shapez
compute quat all property/atom quatw quati &
                 quatj quatk

dump positions all custom 50 out.dump &
      id type x y z &
      c_quat[1] c_quat[2] c_quat[3] c_quat[4] & 
      c_shape[1] c_shape[2] c_shape[3]

timestep 0.05

fix 1 all nve/asphere

fix 2 all langevin/active ${t} ${t} ${d}  &
        82763871 angmom 3.33 100

compute msd_col all msd


thermo_style custom step c_msd_col[1] &
     c_msd_col[2] c_msd_col[3] &
     c_msd_col[4] ke

thermo 100
run 50000
\end{Verbatim}
Notice only the change in the atom style and the langevin fix (and a different dump with shape and orientation to test the ellipsoids proper dynamics). In this new fix langevin/active, after the angmom option we add the angmom value and the activity (100 in this script). You can use the angmom to control the rotational diffusion effect, which in this case is set to $10/3$, the value for spherical passive particles. Since we are dealing with ellipsoids, the inp.data file changes to,
\begin{Verbatim}[numbers=left,xleftmargin=5mm]
# lammps molecular data

1 atoms

1 atom types

1 ellipsoids

0 200 xlo xhi
0 200 ylo yhi
0 200 zlo zhi


Atoms

1  1 1 1.9099 100 100 100

Ellipsoids

1  1 1 1  -0.654 -0.089 -0.285 0.695
\end{Verbatim}
Here, the third column in the Atoms section, is not the diameter, since the shape is defined in collumns 2, 3, and 4 of the Ellipsoids section, but a flag indicating that this particles is an ellipsoid. In the Ellipsoids section, the shape is defined by the diameter in x, y, and z (columns 2, 3, and 4), which is a sphere in this example, and then the four last values are the quaternions quatw, quatx, quaty, and quatz.

\begin{figure}[t]
    \centering
    \includegraphics[width=1\columnwidth]{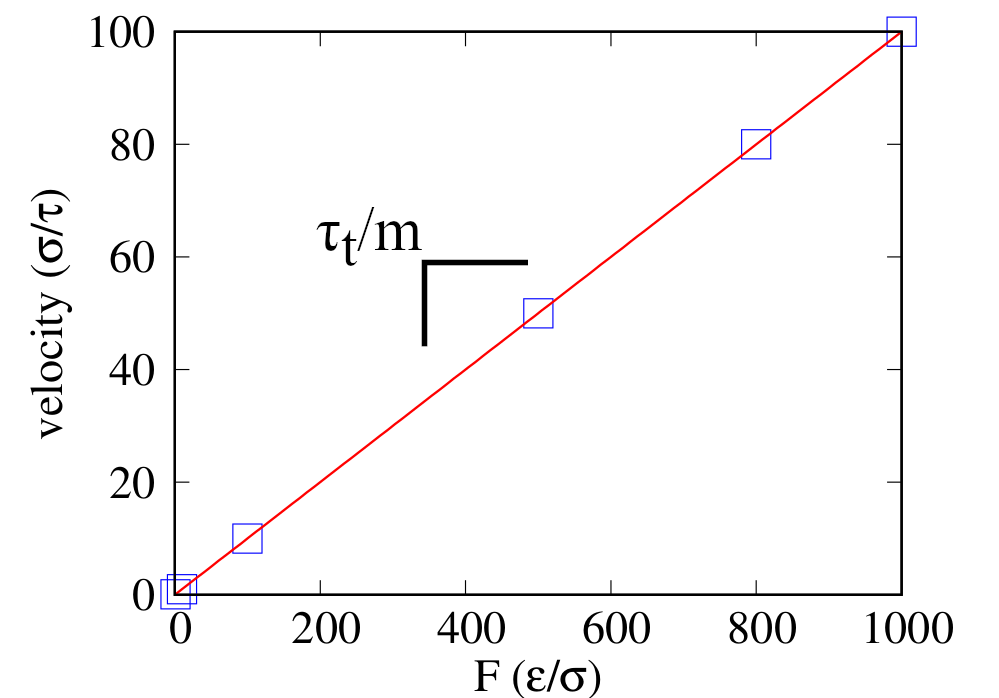}
    \caption{Terminal velocity as a function of the force, for active particles at zero temperature.}
    \label{fig:vforce}
\end{figure}

One simple test to this new active particles class is to measure the velocity at zero temperature for different values of the force. The particles rapidly reach the terminal velocity and, using Eq.~\ref{eq.trans_Langevin_dynamics} at zero temperature, we obtain the relation $v=\frac{\tau_t}{m}F$. This relation is shown in Fig.~\ref{fig:vforce}.

\begin{figure}[t]
    \centering
    \includegraphics[width=1\columnwidth]{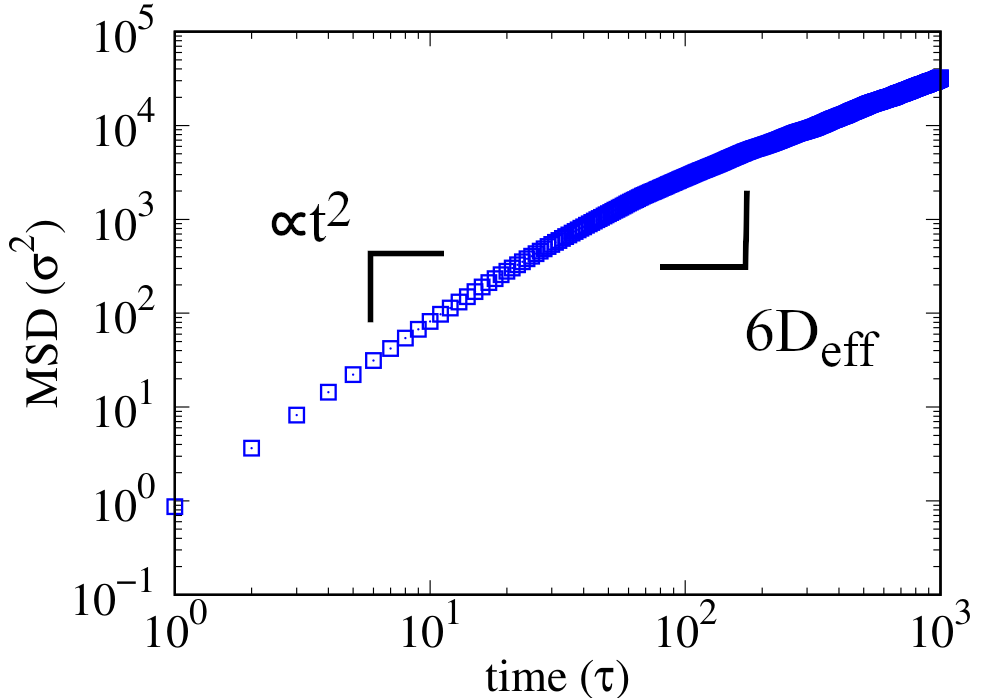}
    \caption{Mean square displacement of an active particle as a function of time., performed for 10000 non-interacting active particles. The two regimes of ballistic and diffusive are seen, where the diffusive regime is linear and proportional to $6D_{eff}$.}
    \label{fig:msd_active}
\end{figure}

Another test is to measure the mean square displacement of an active particle. In Fig.~\ref{fig:msd_active}, we can see the two typical regimes of an active particle, with a ballistic motion at short times and a diffusive motion for longer times. The diffusive motion has an effective diffusion coefficient given by \cite{Bechinger2016},
\begin{equation}
D_{eff}=D_t+\frac{1}{6}\frac{v^2}{D_r},
\end{equation}
where $D_t$ is the translational diffusion coefficient, $v$ the velocity, which we shown in Fig.~\ref{fig:vforce} to be $\tau_t/m$, and $D_r$ the rotational diffusion coefficient. $D_r$ can be related with $D_t$ using  Eqs.~\ref{eq.SE_gen}~and~\ref{eq.SED_gen}, and the scaling factor $\alpha=\tau_t/\tau_r$, which gives the relation,
\begin{equation}
\frac{D_t}{D_r}=\alpha \frac{I}{m},
\end{equation}
 and, for spherical passive particles, recovers Eq.~\ref{eq.coeff_rel}.

\section{Final Discussion}

The LAMMPS package is a very powerful tool. However, like when implementing Molecular Dynamics from scratch, one needs to be very careful with the parametrization of our models. We always need to start with fewer particles and small systems, without parallelization. Then, as done in these notes, perform simple but effective tests to check if we recover the expected physics.

\section{Acknowledgments}

We acknowledge financial support from the Portuguese Foundation for Science and Technology (FCT) under Contracts no. PTDC/FIS-MAC/28146/2017 (LISBOA-01-0145-FEDER-028146), CEECIND/00586/2017, UIDB/00618/2020,  and UIDP/00618/2020. The notes were part of the "Initial Training on Numerical Methods of Active Matter" organized by MSCA-ITN ActiveMatter (Grant: 812780) funded by the EU H2020 program.

\bibliography{LAMMPS.bib}

\end{document}